\documentclass[twocolumn,showpacs,preprintnumbers,amsmath,amssymb,superscriptaddress,floatfix]{revtex4}

\usepackage[dvips]{graphicx}
\usepackage{dcolumn}
\usepackage{bm}
\usepackage{amsmath,amssymb}
\usepackage[titletoc,title]{appendix}
\usepackage{longtable}
\usepackage{bm}
\usepackage{color}

\begin{document}

\setcounter{LTchunksize}{100}
\setlength\LTcapwidth{6.5in}
\setlength\LTleft{0pt}
\setlength\LTright{0pt}

\title{Numerical simulations for the Ising model on three dimensional lattices with coordination number equal 5: static and dynamic critical phenomena.}

\author{Lourdes Bibiana Merino-Sol\'is}
\email{lb.merinosolis@ugto.mx}
\affiliation{%
\ Divisi\'on de Ciencias e Ingenier\'ias,\\
Campus Le\'on de la Universidad de Guanajuato%
}
\author{Francisco Sastre}
\email{sastre@fisica.ugto.mx}
\affiliation{%
Departamento de Ingenier\'ia F\'isica,\ Divisi\'on de Ciencias e Ingenier\'ias,\\
Campus Le\'on de la Universidad de Guanajuato%
}

\date{\today}

\begin{abstract}

In this work we performed numerical simulations for the Ising model on three dimensional lattices with coordination number equal 5.
With Monte Carlo simulations in the static case we evaluated the critical temperature and the static critical exponents $\nu$, $\gamma$ and $\beta$. Once that we have the critical temperature value we investigated the dynamical critical behavior with Glauber dynamics, starting from
disordered states. From our simulations we obtained the values for the dynamic exponent $z=2.037(8)$, the exponent for the autocorrelation $\lambda/z=1.364(5)$, the exponent of the critical initial increase $\theta'=0.109(8)$ and the asymptotic value of the fluctuation-dissipation ratio $X^\infty=0.433(1)$. All of these results are in good agreement to the previous values reported for the 3D Ising model universality class. 
\end{abstract}



\maketitle

\section{Introduction}

\begin{figure}
\begin{center}
\includegraphics[width=0.95\columnwidth]{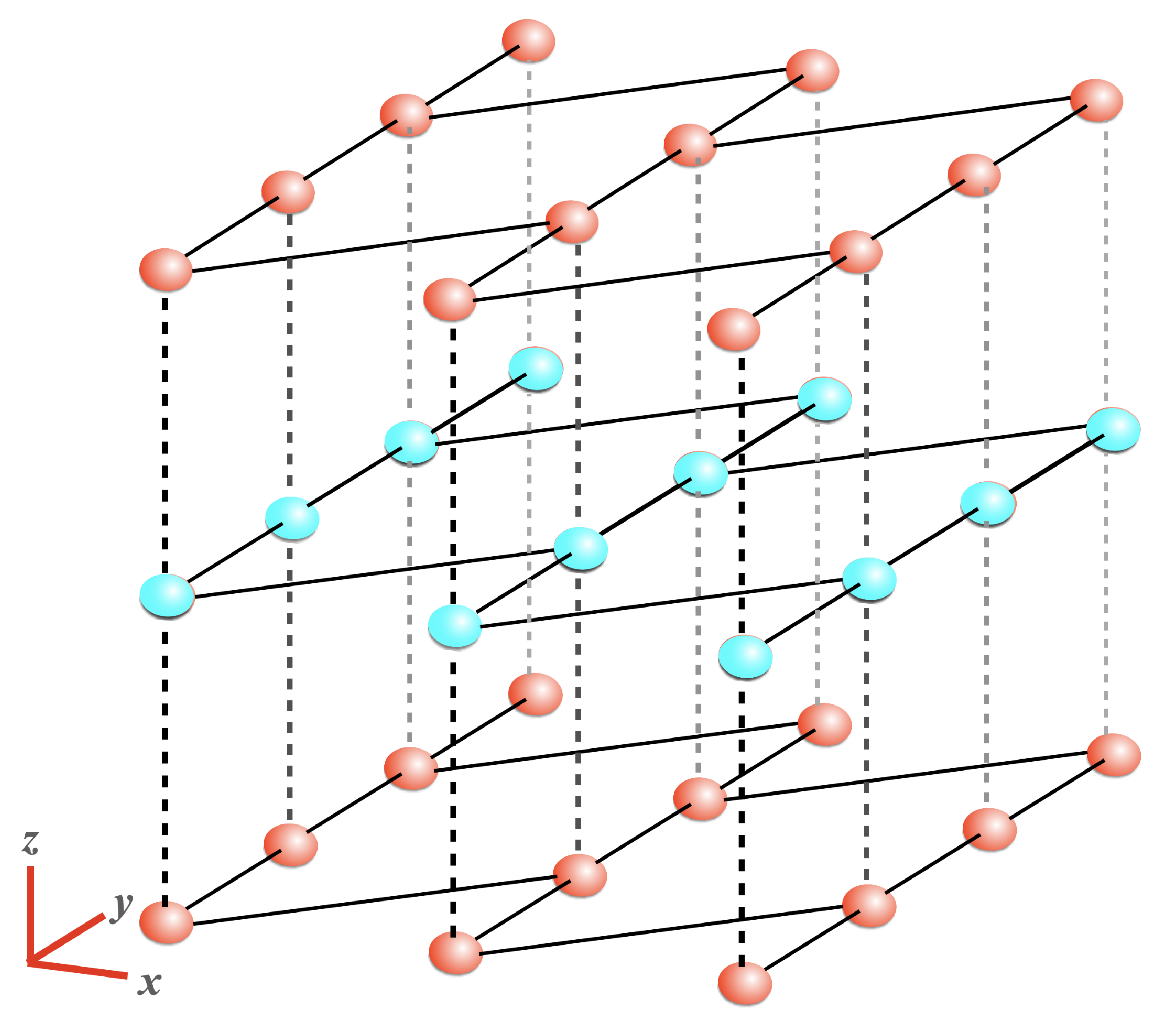}
\caption{\label{redes}
Lattice points in our proposed system. Solid lines show the interaction in the $x$-$y$ plane and the dashed lines show the interaction in the $z$ direction. 
}
\end{center}
\end{figure}

The Ising model is an important tool for the study of phase transitions since can be used to test new numerical and theoretical methodologies.
There is not analytic solution in the three dimensional case,  even though it has been solved in the one and two dimensional cases. 
There have been many numerical studies for the three dimensional simple Ising model on regular lattices both the static critical
phenomena~\cite{Preis2009,Lundow2009,Lundow2018,Hasenbusch2010,Yu2015,Kaupuzs2017,Ferrenberg2018,Sastre2021} and the critical 
dynamic~\cite{Grassberger1995,Jaster1999,Henkel2001,Pleimling2005,Murase2008,Prudnikov2014,Liu2023}. However, to our knowledge, there are
not studies for lattices with even coordination numbers.
In this work we studied the ferromagnetic Ising model on a three dimensional lattice with coordination number (CN) equal to 5. The system used
in this work consists of $L$ two dimensional honeycomb lattices (each layer has $L^2$ sites) stacked as shown in FIG.~\ref{redes}.
In order to simplify the simulations we imposed skew boundary conditions in the $x$ direction and periodic boundary conditions in the other two
directions.
There are works that have used Monte Carlo (MC) simulations on the cubic Ising model to simulate confined
fluids~\cite{Binder1974,Binder1995,Binder2003,Milchev2003,Romero2013,Rodriguez2014}.
We think that the geometry proposed in this work could be useful to study alternative boundary conditions on confined
systems.

In this work we analyzed first the static critical phenomena 
with standard Finite Size Scaling (FSS) theory. The free energy of magnetic systems is described by the scaling ansatz
\begin{equation}
\label{FSS1}
F(L,T)=L^{(2-\alpha)/\nu}{\cal F}(\varepsilon L^{1/\nu},hL^{(\gamma+\beta)/\nu}),
\end{equation}
here $L$ is the linear dimension ($N=L^d$), $\varepsilon=(T-T_c)/T_c$, $h$ is the external magnetic field and $\alpha,~\beta,~\gamma,~\nu$ are the static critical exponents. Our simulations were carried out without external field, then we are setting $h=0$.
From Eq.~(\ref{FSS1}) we can derive the scaling forms for the magnetisation, the susceptibility and the Binder cumulant
\begin{subequations}
\begin{equation}
 M=L^{-\beta/\nu}{\cal M}(\varepsilon L^{1/\nu}), \label{magfss}
\end{equation}
\begin{equation}
 \chi=L^{\gamma/\nu}{\cal X}(\varepsilon L^{1/\nu}), \label{susfss}
\end{equation}
\begin{equation}
 U={\cal U}(\varepsilon L^{1/\nu}). \label{cumfss}
\end{equation}
\end{subequations}
Once that we have the critical temperature we analyzed the relaxation process when our system, initially in a high temperature state, is
suddenly quenched at $t=0$ to the critical temperature $T_c$. For an initial magnetisation $M_0$, the dynamic scaling form for the magnetisation is given by~\cite{Calabrese2006}
\begin{equation}
 M(t,M_0)=t^{-\beta/(\nu z)}{\cal F}_M(t M_0^{1/\kappa}), \label{magt}
\end{equation}
where $z$ is the dynamic critical exponent and the exponent $\kappa$ is related to $z$ with
\begin{equation}
\kappa=\theta'+\beta/(\nu z), \label{kappa}
\end{equation}
here $\theta'$ is the short-time exponent that describes the critical initial increase of the magnetisation, for small values of $M_0$ the expected behavior is~\cite{Janssen1989}
\begin{equation}
 M(t)\propto t^{\theta'}.
\label{crecimiento}
\end{equation}
Starting with $M_0=0$ the autocorrelation function decays with the power law
\begin{equation}
 A(t)=\langle \phi(0,\bm{r})\phi(t,\bm{r})\rangle \propto t^{-\lambda/z}, \label{autot}
\end{equation}
where the autocorrelation exponent $\lambda$ can be expressed in function of $\theta'$ with~\cite{Janssen1992}
\begin{equation}
\lambda=d-z\theta'. \label{dinamicos}
\end{equation}
Other important dynamic quantities analized are the two-time autocorrelation $C(t,s)$ and the autoresponse functions $R(t,s)$ that 
satisfy (see \cite{Henkel2010} and references therein)
\begin{equation}
\begin{aligned}
 C(t,s) {} & =  \langle \phi(s,\bm{r})\phi(t,\bm{r})\rangle -\langle\phi(s,\bm{r})\rangle \langle\phi(t,\bm{r})\rangle \\ 
    & = s^{-b}f_{\text{\tiny $C$}}(t/s),
\label{twocorr}
\end{aligned}
\end{equation}
\begin{equation}
 R(t,s)= \left.\frac{\delta \langle \phi(t,\bm{r})\rangle}{\delta h(s,\bm{r})}\right|_{h=0} = s^{-1-a}f_{\text{\tiny $R$}}(t/s),
\label{tworesp} 
\end{equation}
where $a$ and $b$ are ageing exponents, and $f_{\text{\tiny $C,R$}}(x)$ are scaling functions.
In simple magnetic systems with non-conserved dynamics quenched to $T=T_c$ (our case) we have~\cite{Henkel2003} 
\begin{equation}
a=b=2\frac{\beta}{\nu z}.
\label{identidad} 
\end{equation}
When $x=t/s\gg 1$ the scaling functions obey the power law~\cite{Janssen1989}
\begin{equation}
f_{\text{\tiny $C,R$}}(x) \approx A_{\text{\tiny $C,R$}} x^{-\lambda/z}.
\label{potencialimite} 
\end{equation}

Since the sudden quench to the critical point moves the system away from its equilibrium state,
the correlation and the response no longer satisfy the equilibrium fluctuation-dissipation theorem (FDT): $TR(t,s)=\partial_s C(t,s)$, where
we are setting $k_B=1$.
Cugliandolo and Kurchan~\cite{Cugliandolo1993} proposed that, with the introduction of the fluctuation-dissipation ratio (FDR)
\begin{equation}
 X(t,s):= T \frac{R(t,s)}{\partial_s C(t,s)}
\label{FDR},
\end{equation}
it is possible to characterise when a system is far away from equilibrium.
Godr\`eche and Luck~\cite{Godreche2000-0,Godreche2000} proposed that the asymptotic value of the FDR
\begin{equation}
 X^\infty = \lim_{s\to\infty}\left(\lim_{t\to\infty} X(t,s)\right),
\label{ratio}
\end{equation}
should be a universal quantity in the same way that
the non-equilibrium critical exponents. Combining (\ref{twocorr})-(\ref{ratio}) we have
\begin{equation}
 X^\infty = \frac{\nu z T A_{\text{\tiny $R$}}}{(\nu\lambda-2\beta)A_{\text{\tiny $C$}}},
\label{finalFDR}
\end{equation}
since the partial derivative of the correlation must scale as
\begin{equation}
 \partial_sC(t,s) \approx s^{-1-a}  \frac{(\nu\lambda-2\beta)A_{\text{\tiny $C$}}}{\nu z} x^{-\lambda/z},~\mbox{if}~x\gg1.
\label{finalderivada}
\end{equation}

This work is organized as follow: In Sec,~\ref{teoria} we present the method and the simulation details. In Section~\ref{estatica} we present the numerical results of our simulations in the static phase. While in Section~\ref{dinamica} we present the results of the critical dynamic simulations. Finally in Section~\ref{final} we summarize and give our final conclusions.

\noindent

\section{Model and simulation details}
\label{teoria}
The Hamiltonian for the ferromagnetic Ising model in absence of external magnetic field is given by
\begin{equation}
 {\cal H} = -\sum_{\langle l,m\rangle} \sigma_l\sigma_m,
\label{hamilton}
\end{equation}
where $\sigma_l=\pm 1$, $J>0$ is the coupling constant and the sum runs over all nearest-neighbor pairs. As mentioned in the last section
we will work in a lattices with $N=L^3$ sites. Each spin 
$\sigma_{i,j,k}$ interacts with the spins: $\sigma_{i+1,j,k}$, $\sigma_{i-1,j,k}$, $\sigma_{i,j,k+1}$, $\sigma_{i,j,k-1}$ and
$\sigma_{i+1,j+1,k}$, if $i$ is even, or $\sigma_{i-1,j-1,k}$ if $i$ is odd.
In the static part we will use the dimensionless coupling $K=1/T$, the scaling forms (1) and (2) are still valid with
$\varepsilon=(K-K_c)/K_c$.  
In our simulations each spin will change its sign with probability
\begin{equation}
p(\sigma_l\to -\sigma_l)=\frac{1}{2}[1-\sigma_l\tanh(KH_l)], \label{glauber}
\end{equation}
here $H_j$ is the sum of the 5 spins that interact with the spin $\sigma_l$. 
The instantaneous magnetisation per site is defined as
\begin{equation}
\mu(t)=\frac{1}{N}\sum_k\sigma_k.
\end{equation}
\subsection{Static simulations}
The thermal averages of the magnetisation moments can be evaluate as time averages with
\begin{equation}
\langle M^j\rangle=\frac{1}{\tau-t_o}\sum_{t=t_o}^\tau |\mu(t)|^j, \label{momentos}
\end{equation}
where $t_o$ is the transient time and $\tau-t_o$ is the running time, times are measured in Monte Carlo time steps (MCTS).
Then we can obtain from the simulations the susceptibility
\begin{equation}
\chi =NK\left(\langle M^2\rangle-\langle M\rangle^2\right), \label{susceptibilidad}
\end{equation}
and the Binder cumulant
\begin{equation}
 U=1-\frac{\langle M^4\rangle}{\langle M^2\rangle^2}. \label{cum4}
\end{equation}
The critical coupling $K_c$ and the critical Binder cumulant $U^*$ can be evaluated using the curves $U$ vs. $K$ for different system sizes, since the scaling form (\ref{cumfss})
predicts that, for sufficiently large enough sizes, the curves intersect at the critical point $(K_c,U^*)$ .
The value of the critical Binder cumulant is an universal quantity that depends on cluster shape and boundary
conditions~\cite{Selke2006,Selke2007}. Once that we have the critical point we can evaluate the exponents
$1/\nu$ with the scaling form (\ref{cumfss}). In the same way we can obtain the critical exponents $\beta$ and $\gamma$ with the scaling
forms (\ref{magfss}) and (\ref{susfss}) respectively.

\subsection{Dynamic simulations}
In the critical dynamic we prepared the system flipping spins at random until we reach the desired initial value $M_0$ 
and then we let evolve the system  
following the rule given by (\ref{glauber}).
The time dependent magnetisation after the quench to the critical temperature is evaluated using
\begin{equation}
M(t) = \frac{1}{N} \left\langle \sum_l \sigma_l(t)\right\rangle, 
\label{orderdinamico}
\end{equation}
here $\langle\cdot\rangle$ denotes sampling average.
The autocorrelation can be computed, starting with  $M_0=0$, using
\begin{equation}
A(t) = \frac{1}{N} \left\langle \sum_l \sigma_l(t)\sigma(0)_l\right\rangle. 
\label{autosim}
\end{equation}
The two time correlation function is evaluated using 
\begin{equation}
C(t,s) = \frac{1}{N} \left\langle \sum_l \sigma_l(t)\sigma(s)_l\right\rangle-M(s)M(t). 
\label{corrsim}
\end{equation}
The time derivative $\partial_s C(t,s)$ can be approximate with
\begin{equation}
\begin{aligned}
\partial_s C(t,s) {} & = \frac{1}{N} \left\langle \sum_l \sigma_l(t)[\sigma_l(s+1) -\sigma(s)_l]\right\rangle \\ 
  & -M(t)\Bigl(M(s+1)-M(s)\Bigr). 
\end{aligned}
\label{partsim}
\end{equation}
The response can be evaluated without introducing any magnetic field in the simulation following the approach
proposed by Chatelain~\cite{Chatelain2003,Chatelain2004}.
First we let the system evolve during a time $s$ and then for the next $N$ spin flips attempt (following 
a sequential updating scheme) we store the quantities
\begin{equation}
\Delta_l(s)= \text{\scriptsize $\frac{1}{T}$} \Bigl[\sigma_l(s+1)-\tanh\left(\text{\scriptsize $\frac{1}{T}$} H_l(s)\right)\Bigr],
\label{weiss}
\end{equation}
here $H_l(s)$ is the sum of the 5 spins that interact with the spin $\sigma_l$
at the moment that the spin flip is attempted. 
After the $N$ spin flip attempts we evaluate the two time response function at the time $t>s$ with the relation
\begin{equation}
R(t,s) = \frac{1}{N}\left\langle \sum_l  \sigma_l(t)\Delta_l(s)\right\rangle. 
\label{respsim}
\end{equation}
With this method $C(t,s)$, $R(t,s)$ and $\partial_s C(t,s)$ can be evaluated in the same simulation starting with $M_0=0$.

\begin{figure}
\begin{center}
\includegraphics[width=0.95\columnwidth]{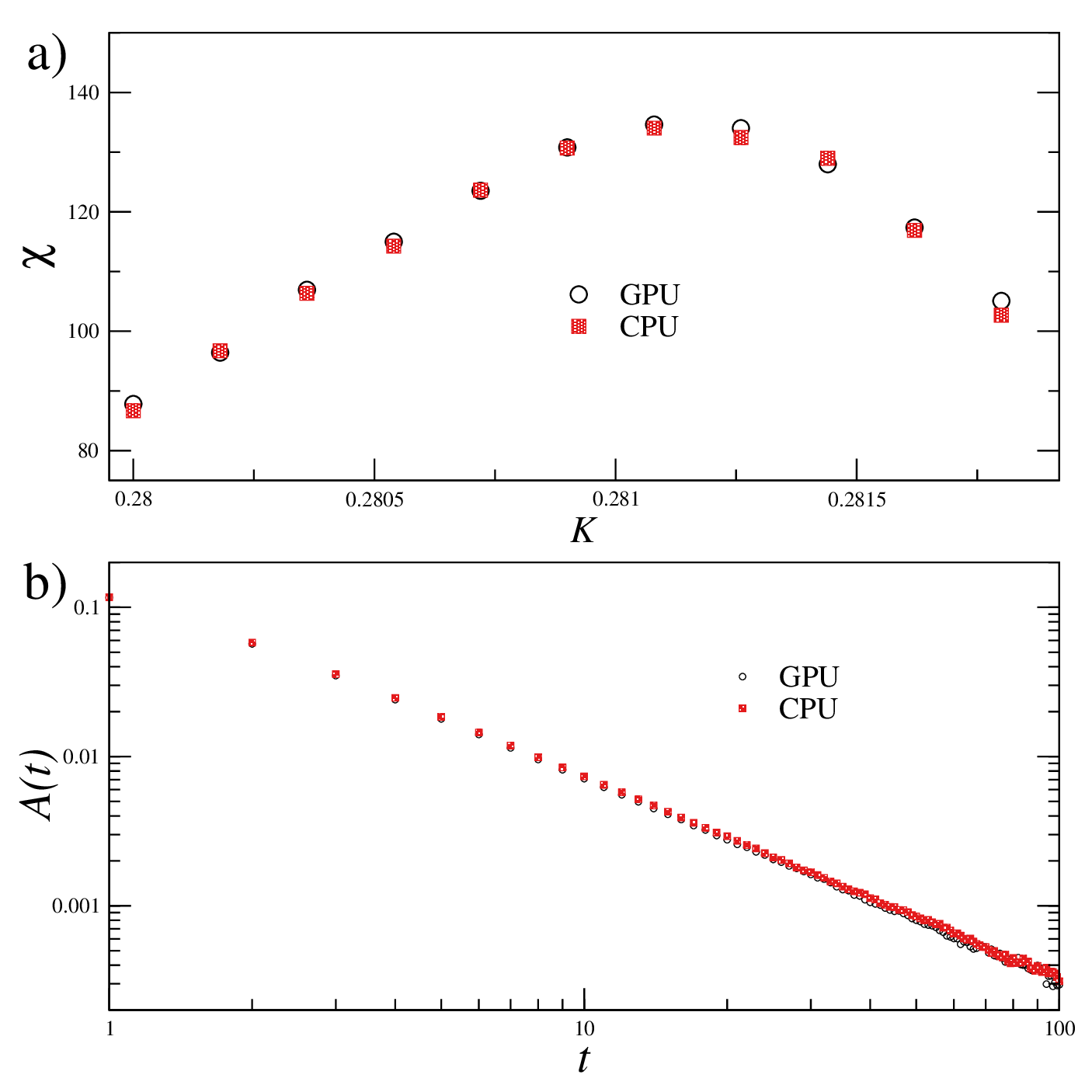}
\caption{\label{comparacion}
a) Numerical simulations for the susceptibility (Eqn. (\ref{susceptibilidad})) in systems of size $L=40$ for GPU, open circles,
and standard CPU simulations using sequential updating, filled squares. b) Dynamic simulations for the autocorrelation 
(Eqn. (\ref{autosim}))  at $T=0.28087$
in systems of size $L=640$ for GPU, open circles, and standard CPU simulations using sequential updating, filled squares.
}
\end{center}
\end{figure}

All the simulations, in the static and dynamic cases, were carried out using the parallel platform CUDA~\cite{CUDA} on a
Graphics Processing Unit (GPU). In the static simulations
we divide the lattice into 4 sub-lattices and performed each MCTS in 4 iterations. For the dynamic simulations we divided the
lattice in 16 sub-lattices and performed each MCTS in 16 iterations. The procedure is equivalent to
update the spins sequentially in a standard Computer Processing Unit (CPU) simulation. In FIG.~\ref{comparacion}) we show, as examples, the
comparison between the GPU and CPU for the a) static and b) dynamic simulations.
 
\section{Static critical phenomena}
\label{estatica}
\begin{figure}
\begin{center}
\includegraphics[width=0.95\columnwidth]{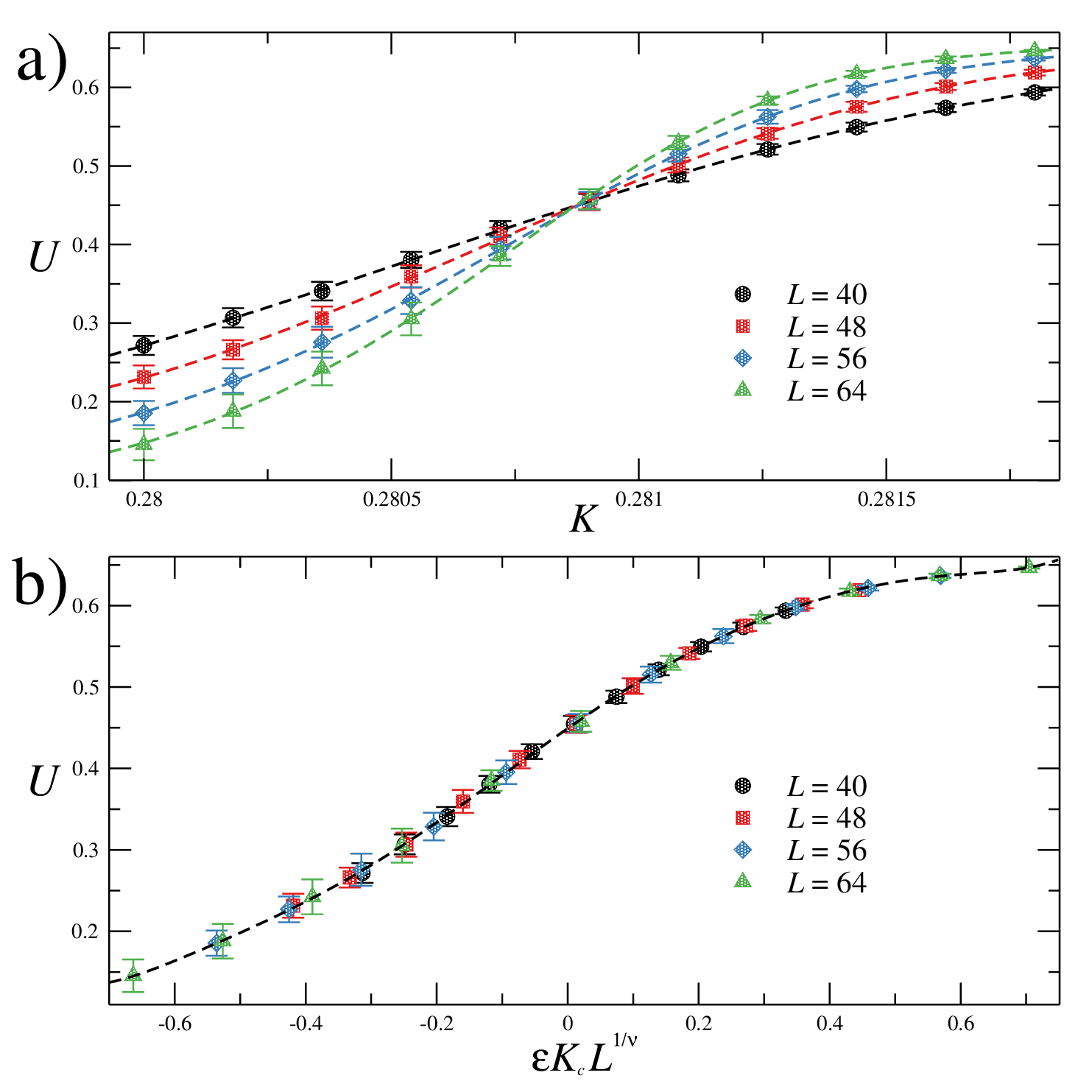}
\caption{\label{cruces}
a) Critical point evaluation from the crossing of the curves $U$ vs. $K$ and b) data collapse from the scaling form~(\ref{cumfss}). In both graphs dots show the numerical results and the dashed lines are numerical fits to the numerical data. From these curves we obtained $K_c=0.280873(13)$,
$\nu=0.627(5)$ and $U^*=0.449(5)$.
}
\end{center}
\end{figure}

The simulations for the evaluation of the critical point and the static critical exponents were performed on lattices of lateral  size
$L=40$, 48, 56 and 64 with transient times from $1\times10^5$ to $1.2\times 10^5$ MCTS and running times from $1\times10^6$ to $1.2\times10^6$
MCTS. Additionally for each set of parameters $K$ and $L$ we performed 30 independent runs. In FIG.~\ref{cruces}a) we show the evaluation of
the critical coupling from the Binder cumulant, we found that all the crossings fall within an interval $\Delta K \sim \times10^{-5}$.
We consider that this range is acceptable as a first estimation of the critical coupling for this geometry. Future works can concentrate 
in the scaling corrections, since there are cases 
where the absence of scaling corrections could occur: like in the two dimensional Ising model on triangular and honeycomb lattices~\cite{Queiroz2000}
or in simulations based on probability distribution functions in the three dimensional Ising model on cubic latices~\cite{Sastre2021}. 
Averaging over the six crossing points we estimate
that the critical coupling is $K_c=0.280873(15)$. This value is greater than the reported values of the three dimensional simple cubic
lattice (CN=6) $K_c=0.221654626(5)$~\cite{Ferrenberg2018},
the bcc lattice (CN=8) $K_c=0.157371(1)$\cite{Lundow2009} 
and the fcc lattice (CN=12) $K_c=0.1020707(2)$\cite{Yu2015}, but smaller that the reported value for  
the diamond lattice (CN=4) $K_c=0.369722(7)$\cite{Lundow2009}. So, our result is consistent with the tendency that the critical
coupling increase as the correlation number decrease.

With our estimated value of the critical coupling we evaluated the critical exponent $\nu$ collapsing the cumulant curves with 
Eq.~(\ref{cumfss}), as shown in FIG~\ref{cruces}b). From our numerical data we obtained the value $\nu=0.627(5)$, which is in good
agreement with the reported value $\nu=0.629912(86)$ from reference~\cite{Ferrenberg2018}. 
As for the critical Binder cumulant we found the value $U^*=0.449(5)$, which is slightly inferior to the reported value $U^*=0.46543(5)$
for the three dimensional Ising model on cubic lattice with periodic boundary conditions~\cite{Ferrenberg2018}.
This was expected, since we are not using periodic boundary conditions in all directions in our simulations.

\begin{figure}
\begin{center}
\includegraphics[width=0.95\columnwidth]{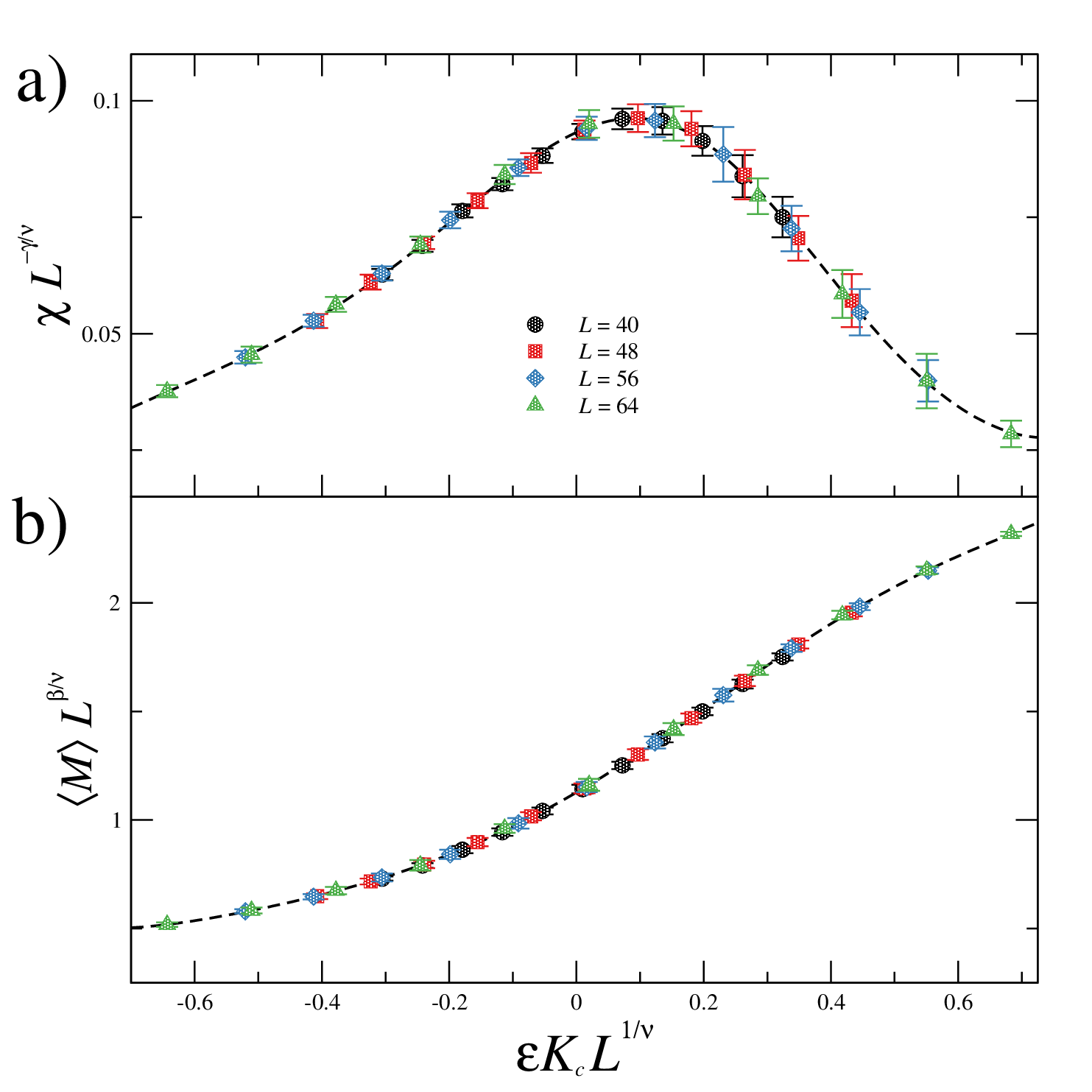}
\caption{\label{universal}
Data collapse for a) the susceptibility, with $\gamma=1.24$ and $\nu=0.63$, and b) the magnetisation, with $\beta=0.32$ and $\nu=0.63$,
for $L=40$ (circles), 48 (squares), 56 (diamonds) and 64 (up triangles).
The dashed lines are numerical fits to the numerical data.
}
\end{center}
\end{figure}

With the estimated values of $K_c$ and $\nu$ we performed FSS for the magnetisation and susceptibility using the scaling forms (\ref{magfss})
and (\ref{susfss}) respectively. From the susceptibility data we found the exponent $\gamma=1.24(1)$ and from the
magnetisation data we found $\beta=0.32(1)$. These results are in good agreement with the reported critical exponents
in Ref.~\cite{Ferrenberg2018} $\gamma=1.23708(33)$ and $\beta=0.32630(22)$.
In FIG.~\ref{universal} we show the data collapse of the susceptibility and
the magnetisation.

We summarized our static results in TABLE~\ref{finalestatico}.

\begin{center}
\begin{table}[ht]
\caption{\label{finalestatico}
Static critical parameters for the three dimensional Ising model with coordination number  5. Results for the cubic
Ising model with periodic boundary conditions from Ref.~\cite{Ferrenberg2018} are listed in the second column.
}
\begin{tabular}{lll}
\hline
\hline
    ~    & ~~~~CN 5~~   &   ~~~~ CN 6~~ \\ 
\hline
  $K_c$~~& 0.280873(13)~~ & 0.221654626(5) \\ 
  $\nu$~~& 0.627(5)    & 0.629912(86)   \\
$\gamma$~~&   1.24(1)   & 1.23708(33)    \\
 $\beta$~~&   0.32(1)   & 0.32630(22)    \\
  $U^*$~~&   0.449(5)  & 0.46543(5)     \\
\hline
\hline
\end{tabular}
\end{table}
\end{center}

\section{Dynamic critical phenomena}
\label{dinamica}

\begin{figure}
\begin{center}
\includegraphics[width=0.95\columnwidth]{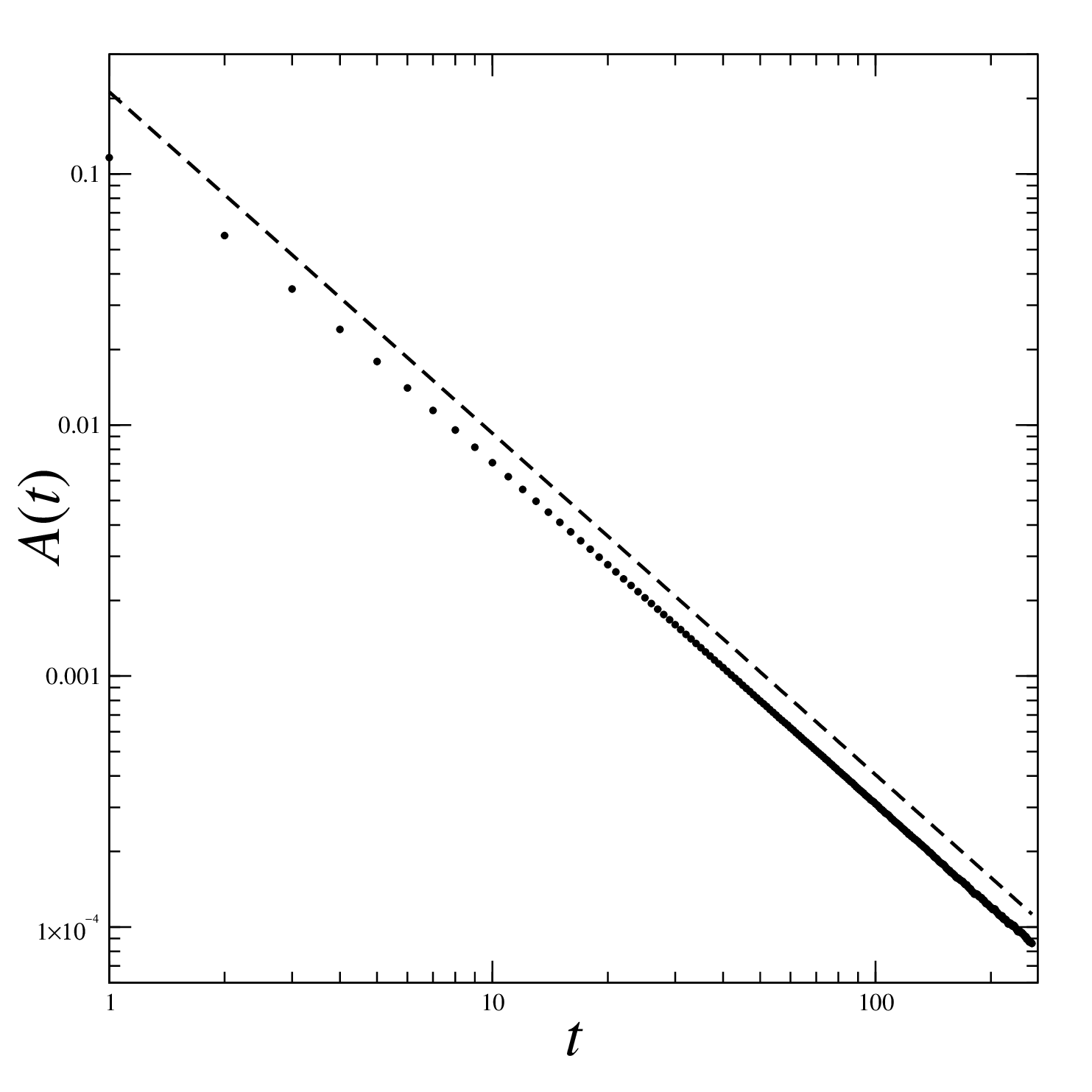}
\caption{\label{lambda}
Autocorrelation as function of time starting with $M_0=0$.
The dashed line shows the expected power law decay (\ref{autot}) with $\lambda/z=1.364$.
}
\end{center}
\end{figure}

For the evaluation of the dynamic exponents $\lambda/z$, $z$ and $\theta'$ we performed simulations with 256 MCTS in lattices of linear
size $L=640$, this assure that there are not size effects in our results. The autocorrelation is evaluated with an initial magnetisation
$M_0=0$, while the time evolution for the magnetisation was measured for $M_0=0.02$, 0.03 and 0.04. In all cases we averaged over 5500 
independent runs. The time evolution of the autocorrelation is show in FIG.~\ref{lambda}, where we observe that the power law decay (\ref{autot}) 
appears after $t_{mic}\sim 10$. With a numerical fit in the time interval $10\le t \le 256$ we obtain the value $\lambda/z=1.364(5)$, that 
is in good agreement with the result
$\lambda/z=1.36$ from Henkel {\it et al.}~\cite{Henkel2001} and
$\lambda/z=1.362(19)$ from Jaster {\it et al.}~\cite{Jaster1999}. 

\begin{figure}
\begin{center}
\includegraphics[width=0.95\columnwidth]{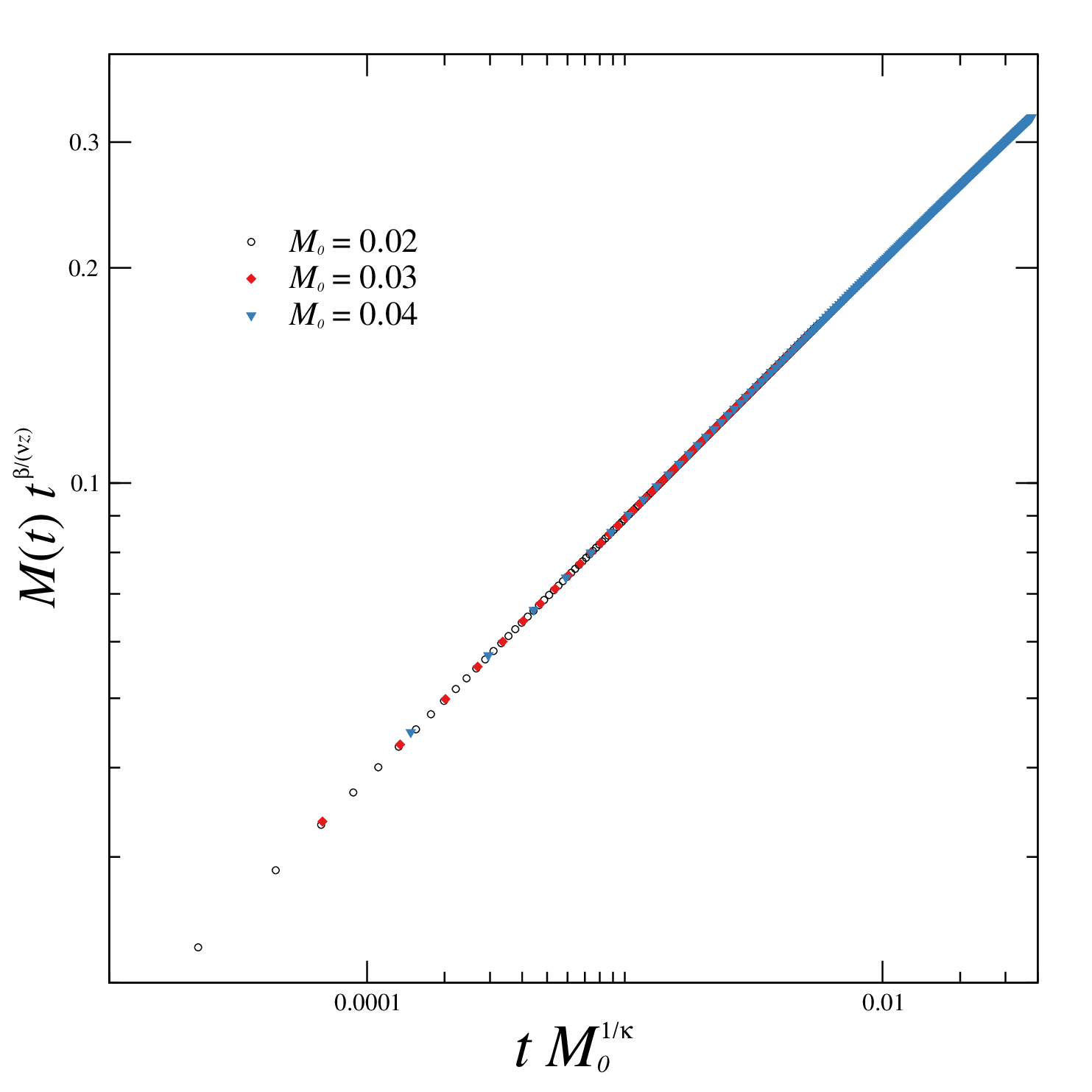}
\caption{\label{colapsotemp}
Data collapse for the $M(t)$ starting from $M_0=0.02$ (circles), $0.03$ (diamonds) and $0.04$ (down triangles) with
$\beta/(\nu z)=0.256$ and $\kappa=0.365$.  
}
\end{center}
\end{figure}

The exponents $z$ and $\kappa$ can be obtained with a data collapse of the magnetisation for the three small non-zero $M_0$ values considered in 
this work. Using the scaling form (\ref{magfss}) we obtain $\beta/(\nu z)= 0.256(3)$ and $\kappa=0.365(2)$ (see FIG.~\ref{colapsotemp}).
Combining (\ref{kappa}) and (\ref{crecimiento}) with $d=3$ we get $z=2.037(8)$ and $\theta'=0.109(8)$. Our dynamical exponent $z$ is
consistent with the recent numerical result $z=2.032(3)$~\cite{Liu2023}, but is slightly larger that the numerical value $z=2.0245(15)$~\cite{Hasenbusch2020} and the five-loop $\epsilon$ expansion value $z=2.0235(8)$~\cite{Adzhemyan2022}.
However, our numerical value for $\beta/(\nu z)$ is in good agreement with the result $\beta/(\nu z)=0.2560(2)$ that is obtained using 
the numerical values of $\beta$, $\nu$ and $z$ from references~\cite{Ferrenberg2018,Adzhemyan2022}.
Our result for the $\theta'$ is also in good agreement with the reported values $\theta'=0.104(3)$ from Grassberger~\cite{Grassberger1995},
$\theta'=0.108(2)$  from Jaster {\it et al.}~\cite{Jaster1999} and
$\theta'=0.108(2)$ from Pleimling and Gambassi~\cite{Pleimling2005} (obtained from the intermediate integrated response).

\begin{figure}[htpb]
\begin{center}
\includegraphics[width=\columnwidth]{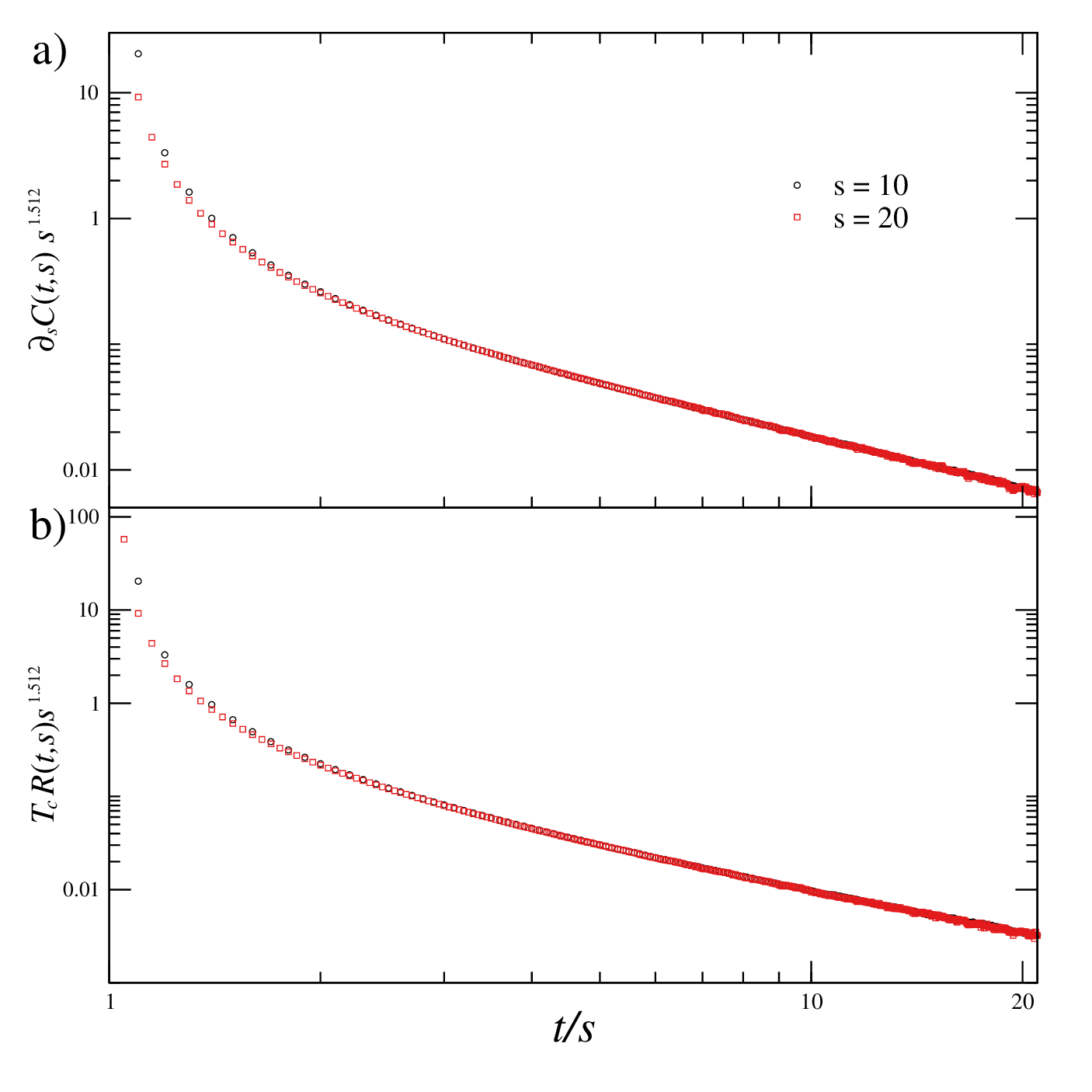}
\caption{\label{dostiempos}
Data collapse for the a)  partial derivative of the correlation and b) response with $2\beta/(\nu z)=0.512$ for $s=10$ (circles) and $s=20$ squares. We observe that an excellent collapse in both cases.
}
\end{center}
\end{figure}

In the simulations of the two time correlation and the response functions we used lattices of size $L=512$ and two waiting times $s=10$ and 20, with 1500
and 2500 independent runs respectively. In FIG.~\ref{dostiempos} we show the scaling for a) the partial derivative of the correlation and b) the response, where we can observe a 
good data collapse using $2\beta/(\nu z) = 0.512$ in the two cases.

\begin{figure}
\begin{center}
\includegraphics[width=\columnwidth]{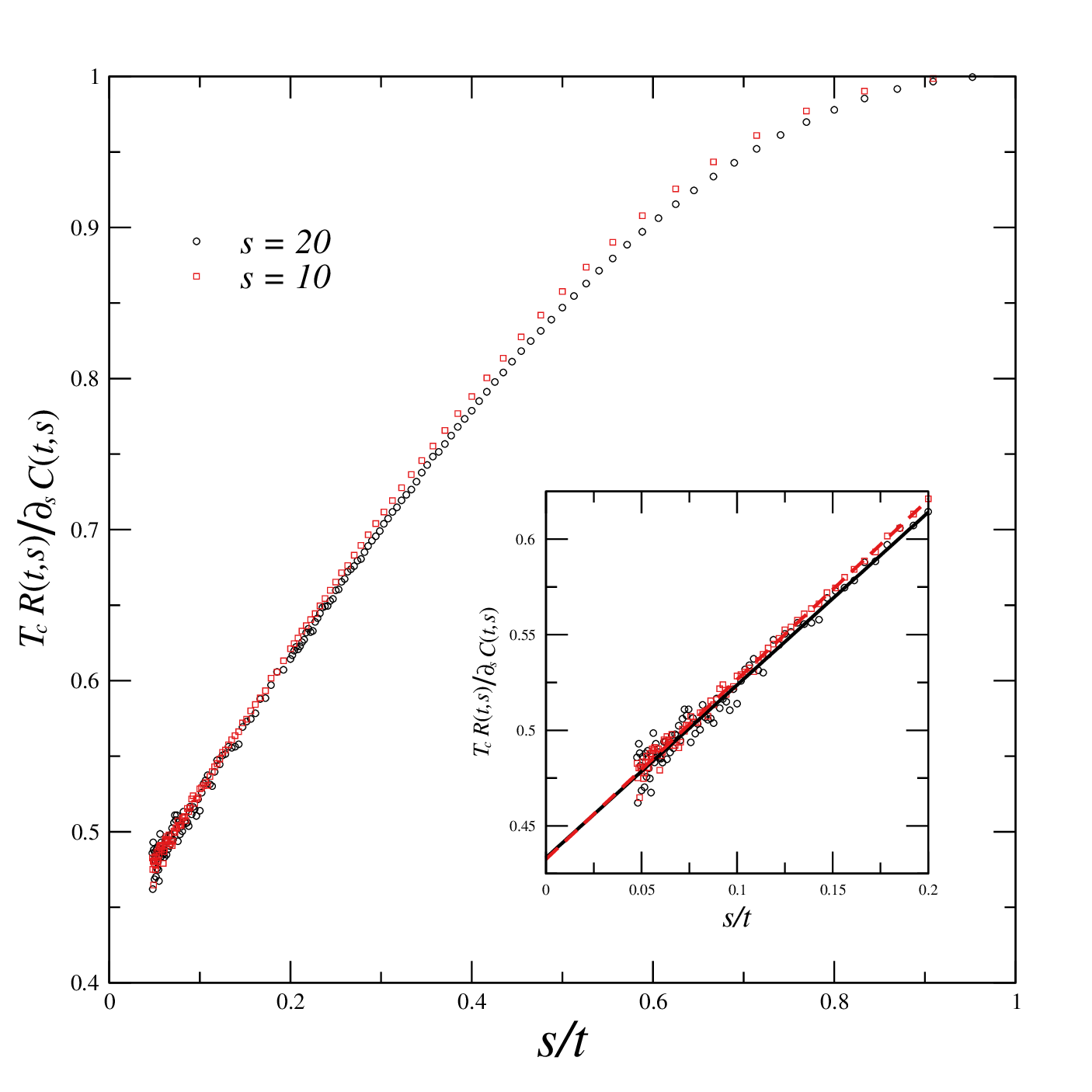}
\caption{\label{razones}
Fluctuation-dissipation ratio $X(t,s)$ as function of $s/t$
for $s=10$ (squares) and $s=20$ (circles).  In the inset we show the evaluation of the asymtotic value of FDR, where the dashed lines
are linear fits to our numerical data. Our estimate is $X^\infty=0.433(1)$.
}
\end{center}
\end{figure}

Finally we evaluated the FDR for the two waiting times considered in this work. In FIG.~\ref{razones} we show $X(t,s)$, Eqn.~(\ref{FDR}),  as function of $s/t$.
From our numerical data we obtained $X^\infty=0.4325(6)$ with $s=10$ and $X^\infty=0.4332(10)$ with $s=20$. Our final estimate is $X^\infty=0.433(1)$, that is in
excellent agreement with the two-loop $\epsilon$ expansion result $X^\infty=0.429(6)$ of Calabrese and Gambassi~\cite{Calabrese2002} and is compatible with the
approximate numerical value $X^\infty\simeq 0.4$ from Godr\`eche and Luck~\cite{Godreche2000}. Our result differ from the numerical result
$X^\infty=0.380(13)$ of Prudnikov {\it et al.}~\cite{Prudnikov2014}, where they used another definition for the
two time correlation function $C(t,s) = \frac{1}{N} \left\langle \sum_l \sigma_l(t)\sigma(s)_l\right\rangle $ instead of (\ref{corrsim}).

We summarized our dynamic results in TABLE~\ref{finaldinamico}.

\begin{center}
\begin{table}[ht]
\caption{\label{finaldinamico}
Dynamic critical parameters for the three dimensional Ising model with coordination number  5. Results for the cubic
Ising model mentioned in the text are listed in the second column.
}
\begin{tabular}{lclcl}
\hline
\hline
    ~  & ~  & ~~CN 5~~ & ~  &   ~~~~ CN 6~~ \\ 
\hline
  $\lambda/z$  & ~   & 1.364(5)& ~   & 1.362(19)~\cite{Jaster1999}, \\ 
 ~& ~          & ~             & ~   & 1.36~\cite{Henkel2001} \\
$\beta/(\nu z)$& ~   & 0.256(3)& ~   & 0.2560(2)~\cite{Ferrenberg2018,Adzhemyan2022}    \\
      $z$~~    & ~   & 2.037(8)& ~   & 2.0245(86)~\cite{Hasenbusch2020},   \\
 ~& ~          & ~             & ~   & 2.0235(8)~\cite{Adzhemyan2022}, \\
 ~& ~          & ~             & ~   & 2.032(3)~\cite{Liu2023} \\
   $\theta'$~~ & ~   & 0.109(8)& ~   & 0.104(3)~\cite{Grassberger1995},   \\
 ~& ~          & ~             & ~   & 0.108(2)~\cite{Jaster1999}, \\
 ~& ~          & ~             & ~   & 0.108(2)~\cite{Pleimling2005} \\
 $X^{\infty}$~ & ~   & 0.433(1)& ~   & 0.429(6)~\cite{Calabrese2002},     \\
 ~& ~          & ~             & ~   & $\simeq 0.4$~\cite{Godreche2000}, \\
 ~& ~          & ~             & ~   & 0.380(13)~\cite{Prudnikov2014} \\
\hline
\hline
\end{tabular}
\end{table}
\end{center}

\section{Conclusions}
\label{final}
We have performed numerical simulations using the parallel platform CUDA to study the static critical phenomena and the critical dynamic in the three dimensional Ising model on an alternate geometry with coordination number equal 5. We present numerical evidence that the static critical exponents $\nu,~\gamma$ and $\beta$, and the dynamic critical exponents  $\lambda/z,~\theta'$ and $z$
are in good agreement with the expected values of the three dimensional Ising model universality class. Our result for the critical Binder
cumulant seems to corroborate that this quantity depends on the boundary conditions used in the simulation. Additional simulations with alternative
boundary conditions in this system could give additional evidence in this matter.
We were able to succesfully implent the method proposed by Chatelain for the evaluation of the response function using the parallel platform CUDA. 
Finally we provided numerical evidence that the limiting value of fluctuation-dissipation ratio in the three dimensional model starting from a dissordered state
by Calabrese and Gambassi with the two-loop $\epsilon$ expansion is correct.

\section*{Acknowledgments}
L. B. Merino-Sol\'{\i}s thanks Conahcyt (Mexico) for fellowship support.

\end{document}